\documentclass[aps,prd,superscriptaddress,nofootinbib,floatfix,preprint]{revtex4-1}

\usepackage[dvipdfmx]{graphicx}
\usepackage{mediabb}
\usepackage{amsmath}
\usepackage{bm}
\usepackage[small,bf]{caption}
\usepackage[subrefformat=parens]{subcaption}
\usepackage{comment}

\newcommand{\beq}{\begin{equation}}
\newcommand{\eeq}{\end{equation}}
\newcommand{\beqa}{\begin{eqnarray}}
\newcommand{\eeqa}{\end{eqnarray}}
\newcommand{\bpr}{\begin{problem}}
\newcommand{\epr}{\end{problem}}
\newcommand{\bcent}{\begin{center}}
\newcommand{\ecent}{\end{center}}
\newcommand{\bfig}{\begin{figure}}
\newcommand{\efig}{\end{figure}}
\newcommand{\bpc}{\begin{picture}}
\newcommand{\epc}{\end{picture}}
\newcommand{\barr}{\begin{array}}
\newcommand{\earr}{\end{array}}
\newcommand{\bitm}{\begin{itemize}}
\newcommand{\eitm}{\end{itemize}}
\newcommand{\bright}{\begin{flushright}}
\newcommand{\eright}{\end{flushright}}
\newcommand{\bminip}{\begin{minipage}}
\newcommand{\eminip}{\end{minipage}}
\newcommand{\btab}{\begin{tabular}}
\newcommand{\etab}{\end{tabular}}

\newcommand{\nnb}{\nonumber}

\newcommand{\hiroshima}{Graduate School of Advanced Science and Engineering, Hiroshima University, Kagamiyama, Higashi-Hiroshima 739-8526, Japan}

\newcommand{\om}{\omega}
\newcommand{\vth}{\vartheta}

\newcommand{\la}{\lambda}

\newcommand{\AlignedEqn}[1]{\begin{equation}\begin{aligned}#1\end{aligned}\end{equation}}

\begin{document}
\title{Sensitivity to axion-like particles with a three-beam stimulated resonant photon collider
       around the eV mass range}

\author{Kensuke Homma\footnote{corresponding author}}\affiliation{\hiroshima}
\author{Fumiya Ishibashi}\affiliation{\hiroshima}
\author{Yuri Kirita}\affiliation{\hiroshima}
\author{Takumi Hasada}\affiliation{\hiroshima}

\date{\today}

\begin{abstract}
We propose a three-beam stimulated resonant photon collider with focused laser fields
in order to directly produce an axion-like particle (ALP) with the two beams and to stimulate
its decay by the remaining one. The expected sensitivity around the eV mass range has been evaluated.
The result shows that the sensitivity can reach the ALP-photon coupling down to 
$\mathcal{O}(10^{-14})$~GeV${}^{-1}$ with 1~J class short-pulsed lasers.
\end{abstract}

\maketitle

\section{Introduction}
CP violation is rather naturally expected from the topological nature
of the QCD vacuum, $\theta$-vacuum, which is required, at least, to solve the $U(1)_A$ anomaly.
Nevertheless, the $\theta$-value evaluated from the measurement of the neutron dipole moment 
indicates the CP conserving nature in the QCD sector.
This so-called strong CP problem is one of the most important
problems yet unresolved in the standard model of particle physics. 
Peccei and Quinn advocated the introduction of a new global $U(1)_{PQ}$ symmetry~\cite{PQ}
in order to dynamically cancel out the finite $\theta$-value expected in the QCD sector
with a counter $\theta$-value around which a massive axion appears as a result of
the symmetry breaking. If the PQ-symmetry breaking scale is much higher than that of
the electroweak scale, the coupling of axion to ordinary matter may be feeble.
This invisible axion can thus be a reasonable candidate for dark matter as a byproduct.

In addition to axion, axion-like particle (ALP) not necessarily requiring the linear relation between
mass and coupling such as in the QCD axion scenario~\cite{AXION}, is also important in the context of 
inflation as well as dark matter in the universe.
Among many possible ALPs, the {\it miracle} model~\cite{MIRACLE} which unifies inflaton and dark matter 
within a single ALP attracts laser-based experimental searches, because the preferred ranges of
the ALP mass $(m_a)$ and its coupling to photons $(g/M)$ 
are $0.01 < m_a < 1$ and $g/M \sim 10^{-11}$~GeV${}^{-1}$, respectively,
based on the viable parameter space consistent with the CMB observation.

So far we have advocated a method to directly produce 
axion-like particles and simultaneously stimulate their decays 
by combining two-color laser fields in collinearly focused geometry~\cite{DEptp}. 
This quasi-parallel photon-photon collision system has been
dedicated to sub-eV axion mass window and the searches have been
actually performed~\cite{PTEP2014,PTEP2015,PTEP2020,SAPPHIRES00,SAPPHIRES01}. 
Given the axion mass window above eV and a typical laser photon energy of $\sim 1$~eV,
stimulated photon-photon collisions with different collision geometry 
has a potential to be sensitive to a higher mass window.
In addition to the well-known axion helio- and halo-scopes, the proposed method can cooperatively provide
unique test grounds totally independent of any of implicit theoretical assumptions on the axion flux
in the Sun as well as in the universe. Therefore, if any of the helio- or halo-scopes detects a hint 
on an ALP, this method can unveil the nature of the ALP via the direct production
and its stimulated decay in laboratory-based experiments by tuning the sensitive mass range
to that specific mass window. In this sense it is indispensable for us
to prepare the independent method for expanding its sensitive mass window as wide as possible.

In this paper we propose a three-beam laser collider
and discuss its expected sensitivity to an unexplored domain for the {\it miracle} model
as well as the benchmark models of the QCD axion based on a realistic set of beam parameters available 
at world-wide high-intensity laser systems.

\section{Formulation dedicated for a stimulated three beam collider}
\begin{figure}[!h]
\begin{center}
\includegraphics[scale=0.65]{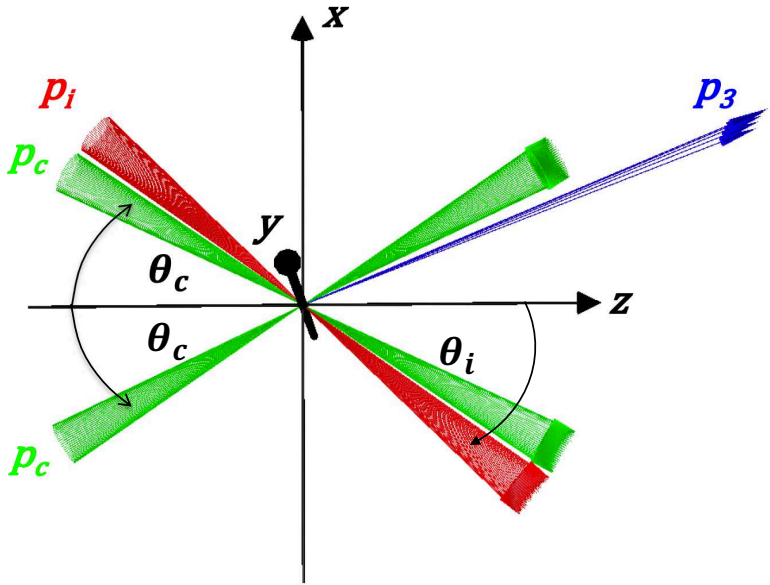}
\end{center}
\caption{Schematic view of a three-beam stimulated photon collider.}
\label{Fig1}
\end{figure}

We focus on the following effective Lagrangian describing
the interaction of an ALP as a pseudoscalar field $\phi_a$ with two photons
\beq\label{Lagrangian}
-{\cal L} = \frac{1}{4}\frac{g}{M}F_{\mu\nu}\tilde{F}^{\mu\nu}\phi_a.
\eeq
As illustrated in Fig.\ref{Fig1}, in the averaged or approximated sense, 
we consider a coplanar scattering with four-momenta $p_i (i=1-4)$
\beq\label{eq_scattering}
<p_c(p_1)> + <p_c(p_2)> \rightarrow p_3 + <p_i(p_4)>,
\eeq
where two focused laser beams, $<p_c>$, create an ALP
with a symmetric incident angle $\theta_c$ and the produced ALP simultaneously
decays into two photons due to an inducing laser beam in the background, $<p_i>$, incident
with a different angle $\theta_i$. As a result of the stimulated decay, emission of
a signal photon $p_3$, is induced. $< >$ symbols reflect the fact that all three beams
contain energy and momentum (angle) spreads at around the focal point. 
The energy uncertainty is caused by Fourier limited short pulsed lasers 
such as femtosecond lasers with the optical frequency, 
while the momentum uncertainty, fluctuations on angle of incidence, is unavoidable 
due to focused fields.
Thus $p_1$, $p_2$ and $p_4$ must be stochastically selected from individual beams,
while $p_3$ is generated as a result of energy-momentum conservation via
$p_1 + p_2 \rightarrow p_3 + p_4$.

We then assume a search for ALPs by scanning $\theta_c$, 
equal incident angles of the two creation beam axes, 
to look for an enhancement of the interaction rate when the resonant condition
\beq
m_a = E_{cms} = 2 \omega_c \sin \theta_c
\eeq
is satisfied, where $m_a$ is the ALP mass, $\omega_c$ is
the central value of single photon energies in the incident creation laser beams
and $E_{cms}$ is center-of-mass system (cms) collision energy between two incident photons.
Because individual incident photons fluctuate around the average beam energy $\omega_c$
and also around the average incident angle $\theta_c$,
this resonance condition has to be evaluated via weighted integral (averaging) over
proper fluctuation distributions as we discuss below.
In the following subsections we thus review necessary formulae
to numerically evaluate the interaction rate by taking generic collision geometry
with asymmetric incident energies and asymmetric incident angles into account.
This asymmetric treatment is essentially required, because unless we implement
the degrees of the spreads at fixed $\theta_c$ and $\omega_c$ depending on experimental parameters,
we cannot determine reasonable discretized steps for the scanning 
over the ALP mass range of interest.

In our previous work~\cite{JHEP} we introduced a theoretical interface allowing the
asymmetric treatment in the case where a single focused beam is used for creation
of an ALP resonance state and the other focused beam sharing the same optical axis
as the creation beam is co-moving for inducing the decay. However, if the sensitive mass range
must be increased, we have to introduce two separated incident beams for the creation part. 
Thus, a modified geometrical treatment for the three separated beams must be reconsidered. 
In \cite{JHEP} we provided formulae only for the case of the scalar field exchange. 
In order to discuss ALPs, we have further extended the formulae to the pseudoscalar exchange case 
with the proper treatment of polarizations affecting the vertex factors~\cite{UNIVERSE00}.
In the following subsections we will provide necessary formulae developed
in \cite{JHEP} and \cite{UNIVERSE00} with necessary modifications for the purpose of this paper.

\subsection{Expression for signal yield in stimulated resonant scattering}
\begin{figure}[!h]
\begin{center}
\includegraphics[scale=0.65]{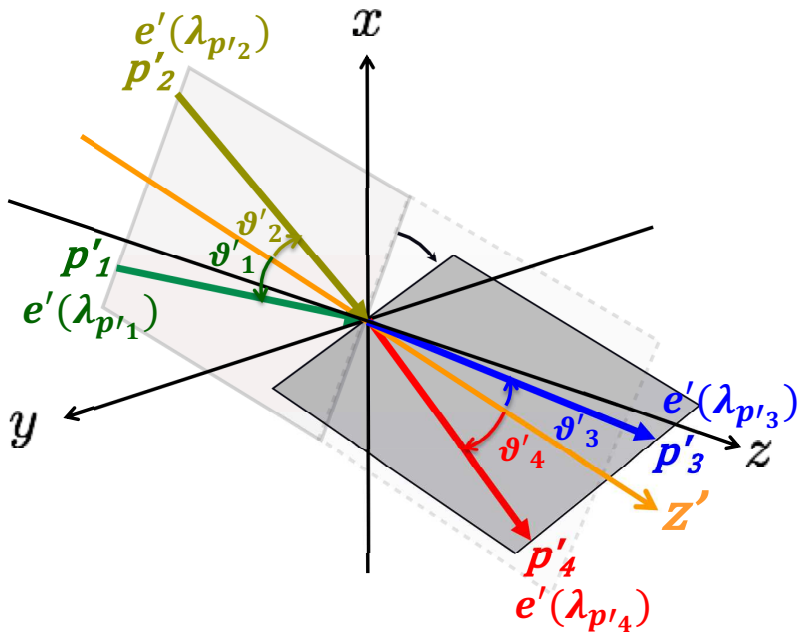}
\end{center}
\caption{Relation between theoretical coordinates with the primed symbol and laboratory coordinates
to which laser beams are physically mapped. The $z^{'}$-axis is theoretically obtainable
so that stochastically selected two incident photons satisfying the resonance condition
have zero pair transverse momentum ($p_T$) with respect to $z^{'}$.
The Lorentz invariant scattering amplitude is calculated on the primed coordinates 
where rotation symmetries of the initial and final state reaction planes around $z^{'}$ are maintained.
Definitions of four-momentum vectors $p'_i$ and four-polarization vectors $e'(\lambda_{p'_i})$
with polarization states $\lambda_{p'_i}$ for the initial state ($i=1,2$) and
final state ($i=3,4$) plane waves are given.
This figure is extracted from~\cite{UNIVERSE00}.
}
\label{Fig2}
\end{figure}

Figure \ref{Fig2} explains the relation between theoretical coordinates 
with the primed symbol and laboratory coordinates
to which laser beams are physically mapped. The $z^{'}$-axis is theoretically obtainable
so that stochastically selected two incident photons satisfying the resonance condition
have zero pair transverse momentum ($p_T$) with respect to $z^{'}$.
The Lorentz invariant scattering amplitude is calculated on the primed coordinates 
where rotation symmetries of the initial and final state reaction planes around $z^{'}$ are maintained.
Definitions of four-momentum vectors $p'_i$ and four-polarization vectors $e'(\lambda_{p'_i})$
with polarization states $\lambda_{p'_i}$ for the initial-state ($i=1,2$) and
final-state ($i=3,4$) plane waves are given.
The conversion between the two coordinates is possible via a simple rotation $\mathcal{R}$
as explained below.
In the following, unless confusion is expected, the prime symbol associated 
with the momentum vectors will be omitted.

We start by reviewing a spontaneous yield of the signal $p_3$, $\mathcal{Y}$,
in the scattering process $p_1 + p_2 \rightarrow p_3 + p_4$
only with two incident photon beams having densities $\rho_1$ and $\rho_2$.
The concept of {\it cross section} is useful for fixed $p_1$ and $p_2$ beams.
In a situation where $p_1$ and $p_2$ largely fluctuate within beams, however,
its convenience is lost.
Thus we apply the following factorization of
{\it volume-wise interaction rate} $\overline{\Sigma}$~\cite{BJ}
instead of {\it cross section} with units of length $L$ and time $s$ in $[\quad]$
\beqa\label{eq_Y}
{\mathcal Y} = N_1 N_2
\left(
\int dt d\bm{r} \rho_1(\bm{r},t) \rho_2(\bm{r},t)
\right)
\times \mbox{\hspace{2cm}} \\ \nnb
\left(
\int dQ W(Q)
\frac{c}{2\om_1 2\om_2} |{\mathcal M}_s(Q^{'})|^2 dL^{'}_{ips}
\right)
\mbox{\hspace{0.2cm}} \nnb\\
\equiv N_1 N_2 {\mathcal D}\left[s/L^3\right] \overline{\Sigma}\left[L^3/s\right]
\mbox{\hspace{3.3cm}}
\eeqa
where the probability density of cms-energy, $W(Q)$, is multiplied for averaging over
the possible range. 
$W(Q)$ is a function of the combinations of photon energies($\omega_{\alpha}$),
polar($\Theta_{\alpha}$) and azimuthal($\Phi_{\alpha}$) angles in laboratory coordinates,
denoted as
\beq\label{eq_QdQ}
Q \equiv \{\omega_{\alpha}, \Theta_{\alpha}, \Phi_{\alpha}\} \quad \mbox{and} \quad
dQ \equiv \Pi_{\alpha} d\omega_{\alpha} d\Theta_{\alpha} d\Phi_{\alpha}
\eeq
for the incident beams $\alpha=1, 2$.
The integral with the weight of $W(Q)$ implements the resonance enhancement
by including the off-shell part as well as the pole in the s-channel amplitude
including the Breit-Wigner resonance function~\cite{JHEP,UNIVERSE00}.

As illustrated in Fig.\ref{Fig2}, $Q^{'} \equiv \{\omega_{\alpha}, \vth_{\alpha}, \phi_{\alpha}\}$
are kinematical parameters in a rotated coordinates $Q^{'}$ constructed from
a pair of two incident wave vectors so that the transverse momentum of the pair
with respect to a $z^{'}$-axis becomes zero. 
The primed coordinates are convenient because the axial symmetry around the $z^{'}$-axis
allows simpler calculations for the following solid angle integral.
The conversions from $Q$ to $Q^{'}$ are thus expressed as rotation matrices on polar and
azimuthal angles:
$\vth_{\alpha} \equiv {\cal R}_{\vth_{\alpha}}(Q)$
and
$\phi_{\alpha} \equiv {\cal R}_{\phi_{\alpha}}(Q)$.

By adding an inducing beam with the central four-momentum $p_4$ having normalized 
density $\rho_4$ with the average number of photons $N_4$,
we extend the {\it spontaneous} yield to the {\it induced} yield, ${\cal Y}_{c+i}$,
with the following extended set of kinematical parameters,
\beq\label{eq_QI}
Q_I \equiv \{Q, \omega_4, \Theta_4, \Phi_4\} \quad \mbox{and} \quad
dQ_I \equiv dQd\omega_4 d\Theta_4 d\Phi_4.
\eeq
as follows
\beqa\label{eq_YI}
{\mathcal Y}_{c+i} = N_1 N_2 N_4
\left(
\int dt d\bm{r} \rho_1(\bm{r},t) \rho_2(\bm{r},t) \rho_4(\bm{r},t) V_4
\right) \times \mbox{\hspace{0.7cm}} \\ \nnb
\left(
\int dQ_I W(Q_I)
\frac{c}{2\om_1 2\om_2} |{\mathcal M}_s(Q^{'})|^2 dL^{'I}_{ips} \mbox{\hspace{0.1cm}}
\right)
\\ \nnb
\equiv  N_1 N_2 N_4 {\mathcal D}_{three}\left[s/L^3\right] \overline{\Sigma}_I\left[L^3/s\right],
\mbox{\hspace{3.1cm}}
\eeqa
where the factor $\rho_4(\bm{r},t) V_4$ is a probability
corresponding to a degree of spacetime overlap of the $p_1$ and $p_2$ beams 
with the inducing beam $p_4$ for a given volume of the $p_4$ beam, $V_4$.
$dL^{'I}_{ips}$ describes an inducible phase space in which the
solid angles of $p_3$ balance solid angles of $p_4$
via energy-momentum conservation within the distribution
of the given inducing beam after conversion from $p_4$ in the primed coordinate system to
the corresponding laboratory coordinate where laser beams are physically mapped. 
With Gaussian distributions $G$,
$W(Q_I)$ is explicitly defined as
\beq\label{eq_WQI}
W(Q_I) \equiv \Pi_\beta G_E(\omega_{\beta}) G_p(\Theta_{\beta},\Phi_{\beta})
\eeq
over $\beta = 1, 2, 4$, where
$G_E$ reflecting an energy spread via Fourier transform limited duration of a short pulse 
and $G_p$ in the momentum space, equivalently the polar angle distribution, are 
introduced based on the properties of
a focused coherent electromagnetic field with an axial symmetric nature for an
azimuthal angle $\Phi$ around the optical axis of a focused beam, as we discuss below.

\subsubsection{\bf Evaluation of spacetime overlapping factor $\mathcal{D}_{three}$ with three beams}\label{ss_Gaussian}
\begin{figure}[!h]
\begin{center}
\includegraphics[scale=0.65]{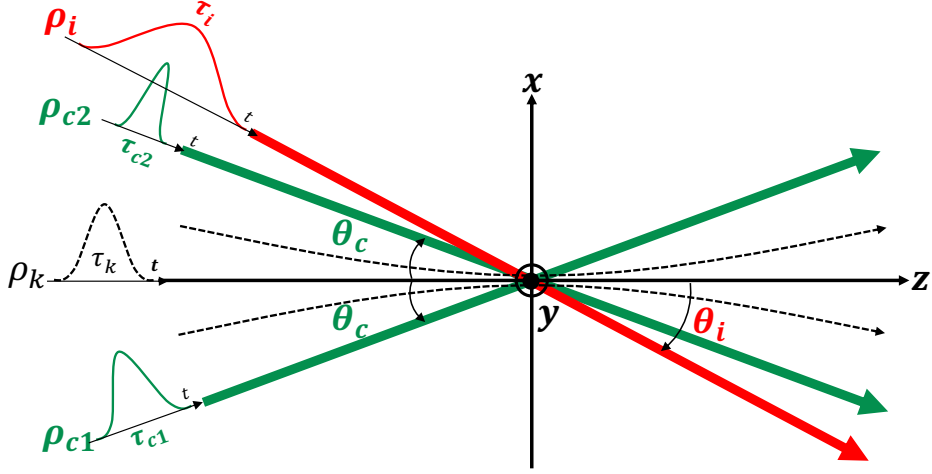}
\end{center}
\caption{Collision geometry between three short pulsed laser beams to define 
the spacetime overlapping factor ${\mathcal D}_{three}$}.
\label{Fig3}
\end{figure}
The factor ${\mathcal D}_{three}$ in Eq.(\ref{eq_YI}) expresses
a spatiotemporal overlapping factor of the focused creation beams (subscript $c1$ and $c2$)
with the focused inducing beam (subscript $i$) in laboratory coordinates.
The following photon number densities $\rho_{k=c1,c2,i}$ deduced from
the electromagnetic field amplitudes based on the Gaussian beam parameterization~\cite{Yariv}
corresponding to the black pulse in Fig.\ref{Fig3} are integrated over spacetime $(t, \bm{r})$:
\beqa\label{eq_rho}
\rho_{k}(t,\bm{r})
 = \left( \frac{2}{\pi} \right)^{\frac{3}{2}}\frac{1}{w_{k}^{2}(ct)c\tau_{k}}
\times \mbox{\hspace{3.4cm}} \\ \nnb
\exp\left( -2\frac{x^{2}+y^{2}}{w_{k}^{2}(ct)} \right)
\exp\left( -2\left( \frac{z-ct}{c\tau_{k}} \right)^{2} \right),
\eeqa
where $w_k$ are the beam radii as a function of time $t$ whose origin is 
set at the moment when all the pulses reach the focal point, 
and $\tau_k$ are the time durations of the pulsed laser beams
with the speed of light $c$ and the volume for the inducing beam $V_i$ is defined as
\beq
V_i = (\pi/2)^{3/2} w^2_{i0} c\tau_i ,
\eeq
where $w_{i0}$ is the beam waist (minimum radius) of the inducing beam.
As a conservative evaluation, the integrated range for the overlapping factor is
limited in the Rayleigh length 
\beq
{z_i}_R = \frac{\pi {w_i}^2_0}{\lambda_i}
\eeq
with the wavelength of the inducing beam $\lambda_i$
only around the focal point where the induced scattering probability is maximized.

Figure \ref{Fig3} illustrates spacetime pulse functions propagating
along individual optical axes of the three beams 
which are defined by rotating coordinates in Eq.(\ref{eq_rho}) around $y$-axis.
$\rho_{c1}$, $\rho_{c2}$ and $\rho_{i}$ are defined with the rotation angles: $\theta_c$,
$-\theta_c$ and $-\theta_i$, respectively. That is, we assume symmetric incident angles
between the two creation laser beams and supply the inducing laser so that photon four-momenta
satisfy energy-momentum conservation with respect to a fixed central value for signal photon four-momenta.

The overlapping factor with units of [$s/L^{3}$]
can be analytically integrated over spatial coordinates
and is eventually obtained by numerically integrating over time from $-z_{iR}/c$ to $0$
as follows:
\begin{equation}
\label{Dfactor_3beam}
\begin{split}
        D_{three}\,[s/L^3] =
                & \left(\frac{1}{\pi}\right)^\frac{3}{2}
                        2^3 w^2_{i0}
                        \int^{0}_{-\frac{z_{iR}}{c}}dt
                        \frac{1}{w^3_c w_i c^2 \tau_{c1}\tau_{c2}}\sqrt{\frac{1}{2\left(2w^2_i + w^2_c\right)}}
                        \\
                & \sqrt{\frac{J}{HS}}
                        \exp\left[\frac{T^2}{4S} - R\right]
                        \exp\left[-2\frac{\tau^2_{c2}\tau^2_i + \tau^2_{c1}\tau^2_i + \tau^2_{c1} \tau^2_{c2}}{\tau^2_{c1}\tau^2_{c2}\tau^2_{i}}t^2\right].
\end{split}
\end{equation}
The individual variables in Eq.(\ref{Dfactor_3beam}) are summarized as follows,
where
we use abbreviations $\mathbf{C_k} = \cos\theta_k$ and $\mathbf{S_k} = \sin\theta_k$
for $k=c1, c2, i$  and we assume 
$\mathbf{C_c} \equiv \mathbf{C_{c1}} = \mathbf{C_{c2}}$,
$\mathbf{S_c} \equiv \mathbf{S_{c1}} = \mathbf{S_{c2}}$,
$w_c \equiv w_{c1} = w_{c2}$, 
$d_c \equiv d_{c1} = d_{c2}$
and
$f_c \equiv f_{c1} = f_{c2}$, 
because two creation beams are incident with a symmetric angle and 
focused with equal beam diameters and focal lengths.
\AlignedEqn{
        \label{eq:parameters_Dfactor}
        J\,[L^6 \cdot s^4] & = w^2_c w^2_i c^2 \tau^2_{c1} \tau^2_{c2} \tau^2_i,
        & \quad
        H\,[L^4 \cdot s^4] & = 2 \left(2 C \mathbf{C_c}^2 + D \mathbf{C_i}^2 + E \mathbf{S_c}^2 + F \mathbf{S_c}^2 + G \mathbf{S_i}^2\right), \\
        S\,[1/L^2] & = \frac{O}{J} - P,
        & \qquad
        T\,[1/L] & = - \left(N + Q\right), \\
        R\,[1] & = - \frac{4}{HJ} \left(B_c M - B_i G\right)^2.
}
The parameters $B, C, D, E, F$ and $G$ are
\AlignedEqn{
        \label{eq:parameters_Dfactor2}
        B_k\,[L] & = 2 c t \mathbf{S_k},
        & \qquad
        C\,[L^4 \cdot s^4] & = w^2_i c^2 \tau^2_{c1} \tau^2_{c2} \tau^2_i,
        & \qquad
        D\,[L^4 \cdot s^4] & =  w^2_c c^2 \tau^2_{c1} \tau^2_{c2} \tau^2_i,
        \\
        E\,[L^4 \cdot s^4] & = w^2_c w^2_i \tau^2_{c2} \tau^2_i,
        & \qquad
        F\,[L^4 \cdot s^4] & =  w^2_c w^2_i \tau^2_{c1} \tau^2_i,
        & \qquad
        G\,[L^4 \cdot s^4] & =  w^2_c w^2_i \tau^2_{c1} \tau^2_{c2}.
}
The parameters $M, N, O, P$ and $Q$ are
\AlignedEqn{
        \label{eq:parameters_Dfactor3}
        M\,[L^4 \cdot s^4] & = E - F,
        \qquad
        N\,[1/L] = 4t \left(\frac{\tau^2_{c2}\tau^2_{i} \mathbf{C_c} + \tau^2_{c1}\tau^2_{i} \mathbf{C_c} + \tau^2_{c1}\tau^2_{c2}\mathbf{C_i}}{c \tau^2_{c1}\tau^2_{c2}\tau^2_{i}}\right), \\
        O\,[L^4 \cdot s^4] & = 2\left(2 C \mathbf{S_c}^2 + D \mathbf{S_i}^2 + E \mathbf{C_c}^2 + F \mathbf{C_c}^2 + G  \mathbf{C_i}^2\right), \\[2pt]
        P\,[1/L^2] & = \frac{4}{HJ}\left\{\left( \mathbf{C_i} \mathbf{S_i}D + \mathbf{C_c} \mathbf{S_c} M\right)^2 - 2  \mathbf{C_c} \mathbf{S_c} \mathbf{C_i} \mathbf{S_i} M G  +  \mathbf{C_i}^2 \mathbf{S_i}^2 G^2 -2  \mathbf{C_i}^2 \mathbf{S_i}^2 D G\right\},
        \\
        Q\,[1/L] & = \frac{4}{HJ} \left\{\mathbf{C_c} \mathbf{S_c} M \left(B_c M + B_i G\right) - \mathbf{C_i} \mathbf{S_i} \left(B_c M G - B_i G^2 + D B_i G - D B_c M\right)\right\}.
}
The beam parameters relevant to focused geometry used above are expressed as
\AlignedEqn{
        \label{eq:laserparameters}
        w_k & = w_{k0} \sqrt{1 + \frac{c^2 t^2}{{z_{kR}}^2}},
        \qquad
        w_{k0} & = \frac{\lambda_k}{\pi \vartheta_{k0}} ,
        \qquad
        \vartheta_{k0} & = \arctan \left(\frac{d_k}{2f_k} \right)
}
with $k = c1, c2, i$.

\subsubsection{\bf Evaluation of inducible volume-wise interaction rate, $\overline{\Sigma}_I$}
\begin{figure}[!h]
\begin{center}
\includegraphics[scale=0.80]{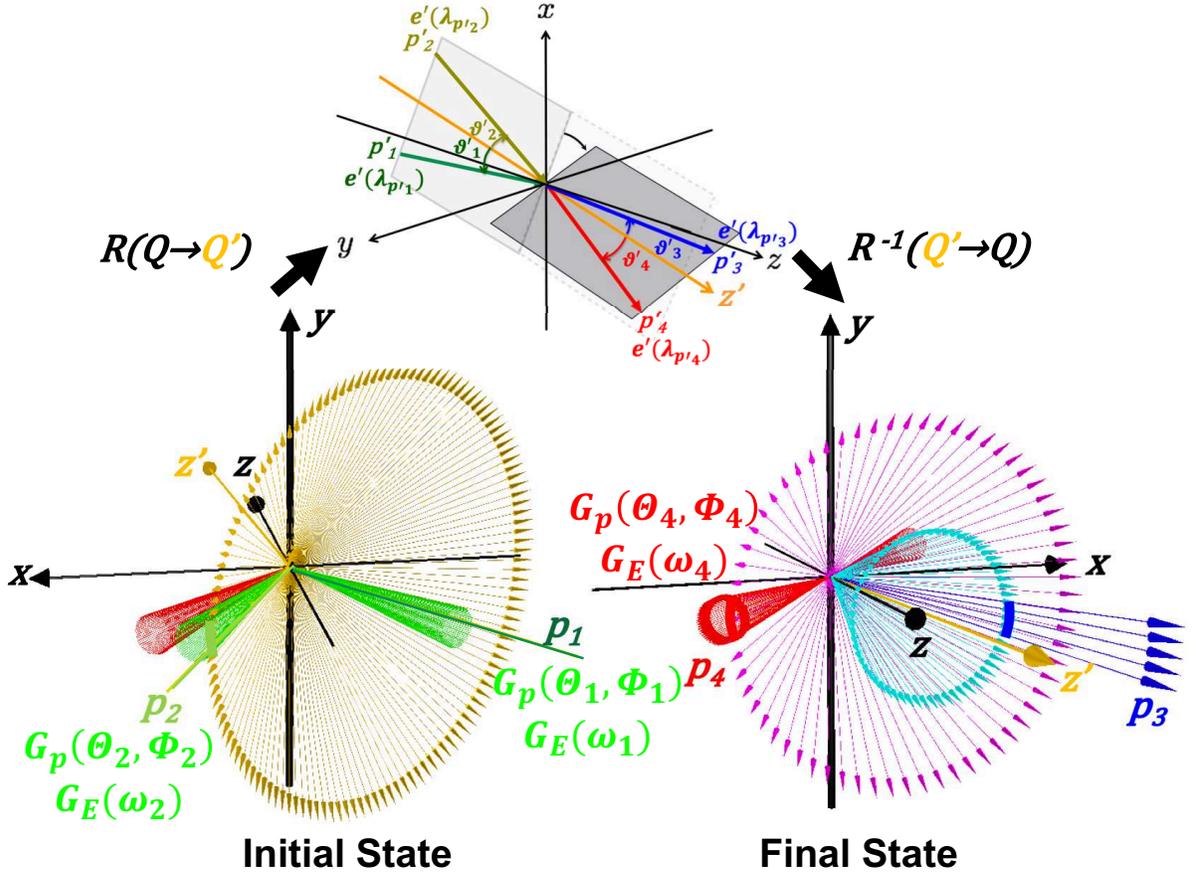}
\end{center}
\caption{Flow of the numerical calculation.
The left figure depicts the initial state of two scattering photons with incidence
of two creation beams (green) and an inducing beam (red),
while the right figure indicates the final state photons, that is, the inducing beam photons
and signal photons (blue) in the laboratory coordinates by omitting the outgoing two creation beams.
The top figure is to remind of the scattering amplitude calculation in the primed coordinates.
Probability distribution functions in momentum space $G_p$ as a function of polar angles $\Theta_i$
and azimuthal angles $\Phi_i$ in the laboratory coordinates and
those in energy $G_E(\omega_i)$  for individual photons $i=1,2,4$ are
assigned to individual focused beams by denoting the normalized Gaussian distributions as $G$.
}
\label{Fig4}
\end{figure}
Performing the analytical integral for $\overline{\Sigma}_I$ in Eq.(\ref{eq_YI})
is not practical and we are forced to evaluate it with the numerical integral.
The expression for $|\mathcal{M}_s(Q^{'})|^2$ in the inducible volume-wise interaction rate is 
fully explained in~\cite{UNIVERSE00}. In this paper we focus on how to implement the numerical integral
configured for a three-beam collider with focused beams.
Figure \ref{Fig4} illustrates the entire flow of the calculation.
The left figure depicts the initial state of two scattering photons with incidence
of two creation beams (green) and an inducing beam (red),
while the right figure indicates the final state photons, that is, the inducing beam photons
and signal photons (blue) in the laboratory coordinates by omitting the outgoing two creation beams.
The top figure is to remind of the scattering amplitude calculation in the primed coordinates.
Probability distribution functions in momentum space $G_p$ as a function of polar angles $\Theta_i$
and azimuthal angles $\Phi_i$ in the laboratory coordinates and 
those in energy $G_E(\omega_i)$  for individual photons $i=1,2,4$ are 
assigned to individual focused beams by denoting the normalized Gaussian distributions as $G$.
The actual steps for the calculations are as follows:
\begin{enumerate}
\item Select a finite-size segment of $p_1$ from 
given $G_E(\omega_1) G_p(\Theta_1,\Phi_1)$ distributions.
\item Find $p_2$ which satisfies the following resonance condition 
\beq
m_a = E_{cms} = 2\sqrt{\omega_1\omega_2}\sin\left(\frac{\vartheta_1+\vartheta_2}{2}\right)
\eeq
with respect to the selected $p_1$
and to a finite energy segment in $G_E(\om_2)$ for a given mass parameter $m_a$.
The possible $p_2$ candidates satisfying the resonance condition form
the yellow thin cone around the $p_1$-axis reflecting the width of the Breit-Wigner function 
as shown in the left figure.
\item Form a $z^{'}$-axis so that the pair transverse momentum, $p_T$, becomes zero,
which is defined as zero-$p_T$ coordinates (primed coordinates)
in contrast to the laboratory coordinates to which the three beams are physically mapped.
Only a portion of the creation beam prepared for $p_2$ overlapping with the yellow cone
can effectively contribute to the resonance production and, hence, the field
weight for the pair can be eventually evaluated by properly reflecting 
$G_E(\omega_1) G_p(\Theta_1,\Phi_1)$ and $G_E(\omega_2) G_p(\Theta_2,\Phi_2)$.
\item Convert the polarization vectors $e_i(\la_i)$ as well as the momentum vectors
from the laboratory coordinates to the zero-$p_T$ coordinates through the coordinate rotation 
$\mathcal{R}(Q \rightarrow Q^{'})$.
\item The axial symmetric nature of possible final-state momenta $p^{'}_3$ and $p^{'}_4$
around $z^{'}$ is represented by the light-blue and magenta vectors in the right figure.
A spontaneous scattering probability with the vertex factors using the primed polarization
vectors in the planes containing the four photon wave vectors is 
calculated in the given zero-$p_T$ coordinates as illustrated in the top figure.
By using the axial symmetric nature around $z^{'}$,
the probability can be integrated over possible final state planes containing $p^{'}_3$ and $p^{'}_4$.
\item In order to estimate the inducing effect for given $G_E(\om_4)G_p(\Theta_4,\Phi_4)$
distributions fixed in the laboratory coordinates, a matching fraction of $p_4$ is calculated
after rotating the primed vectors back to those in the laboratory coordinates
from the zero-$p_T$ coordinates via the inverse rotation $\mathcal{R}^{-1}(Q^{'} \rightarrow Q)$.
Based on the spread of $G_E(\om_4)$, the weights along the overlapping belt between the magenta
and red vectors in the right figure are taken into account as the enhancement factor for
the stimulation of the decay.
\item Due to energy--momentum conservation, $p_3$ must balance with $p_4$.
Thus a signal energy spread via $\om_s \equiv \om_3 = \om_1+\om_2-\om_4$ and also the polar-azimuthal
angle spreads by taking the $G_E(\om_4)G_p(\Theta_4,\Phi_4)$ distributions into account
are automatically determined.
The volume-wise interaction rate $\overline{\Sigma}_I$ is then integrated over the inducible
solid angle of $p_3$ reflecting all the energy and angular spreads included in the focused three beams.
\item With Eq.~(\ref{eq_YI}) the signal yield ${\cal Y}_{c+i}$ can be evaluated.
\end{enumerate}

\section{Expected sensitivity}
\begin{table}[h!]
\caption{Experimental parameters used to numerically calculate the upper limits on the coupling--mass relations.
}
\begin{center}
\begin{tabular}{lr}  \\ \hline
Parameter & Value \\ \hline
Centeral wavelength of creation laser $\lambda_c$ & 800~nm($\omega$)/400~nm($2\omega$)/267~nm(3$\omega$)\\
Relative linewidth of creation laser, $\delta\omega_c/<\omega_c>$ &  $10^{-2}$\\
Duration time of creation laser, $\tau_{c}$ & 30 fs \\
Creation laser energy per $\tau_{c}$, $E_{c}$ & 1~J \\
Number of creation photons($\omega$),  $N_c$ & $4.03 \times 10^{18}$ photons\\
Number of creation photons($2\omega$), $N_c$ & $2.01 \times 10^{18}$ photons\\
Number of creation photons($3\omega$), $N_c$ & $1.34 \times 10^{18}$ photons\\
Beam diameter of creation laser beam, $d_{c}$ & 60~mm\\
Polarization & linear (P-polarized state) \\ \hline
Central wavelength of inducing laser, $\lambda_i$   & 1300~nm\\
Relative linewidth of inducing laser, $\delta\omega_{i}/<\omega_{i}>$ &  $10^{-2}$\\
Duration time of inducing laser beam, $\tau_{i}$ & 100~fs\\
Inducing laser energy per $\tau_{i}$, $E_{i}$ & $0.1$J \\
Number of inducing photons, $N_i$ & $6.54 \times 10^{17}$ photons\\
Beam diameter of inducing laser beam, $d_{i}$ & $30$~mm\\
Polarization & circular (left-handed state) \\ \hline
Focal length of off-axis parabolic mirror, $f_c=f_i$ & 600~mm\\
Overall detection efficiency, $\epsilon$ & 1\% \\
Number of shots, $N_{shots}$   & $10^5$ shots\\
$\delta{N}_{S}$ & 100\\
\hline
\end{tabular}
\end{center}
\label{Tab1}
\end{table}

We evaluate search sensitivities based on the concept of a three-beam stimulated
resonant photon collider (${}^\mathrm{t}$SRPC) with variable incident angles for 
scanning ALP masses around the eV range as illustrated in Fig.\ref{Fig1}. 
By assuming high-intensity femtosecond lasers such as Titanium:Sapphire lasers
with 1~J pulse energy for simplicity, 
we consider that two identical creation beams with the central photon energy $\omega_c$
and the time duration $\tau_c$ are symmetrically incident 
with the same beam incident angle $\theta_c$ and 
an inducing laser with the central photon energy
$\omega_i \equiv u\omega_c$ with $0<u<1$ is incident with the corresponding angle 
which satisfies energy-momentum conservation by requiring a common
signal photon energy $(2-u)\omega_c$ independent of various incident angle combinations.
The central wavelength is around 800~nm and then we assume the ability to 
produce high harmonic waves from the fundamental wavelength
for creation beams and to generate an inducing beam with a non-integer number $u$ 
based on the optical parametric amplification (OPA) technique
in order to discriminate signal waves against the integer number high harmonic waves
originating from the creation beams. 
Since the OPA technique cannot achieve the perfect conversion from the fundamental wavelength, 
we assume 0.1~J pulse energy and also the elongation of the pulse duration for the inducing beam
compared to that in the creation beam.
Table~\ref{Tab1} summarizes assumed parameters for two identical creation laser beams 
and an including laser beams as well as the common focusing and statistical parameters.

Given a set of three-beam laser parameters $P$ in Table \ref{Tab1},
the number of stimulated signal photons, $N_{obs}$, is expressed as
\beq\label{Nobs}
N_{obs} = {\cal Y}_{c+i}(m_a, g/M ; P) N_{shot} \epsilon ,
\eeq
which is a function of ALP mass $m_a$ and coupling $g/M$,
where $N_{shot}$ the number of laser shots and
$\epsilon$ the overall efficiency of detecting $p_3$.
For a set of $m_a$ values with an assumed $N_{obs}$,
a set of coupling $g/M$ can be estimated by numerically solving Eq.(\ref{Nobs}).

\begin{figure}[!h]
\begin{center}
\includegraphics[scale=0.80]{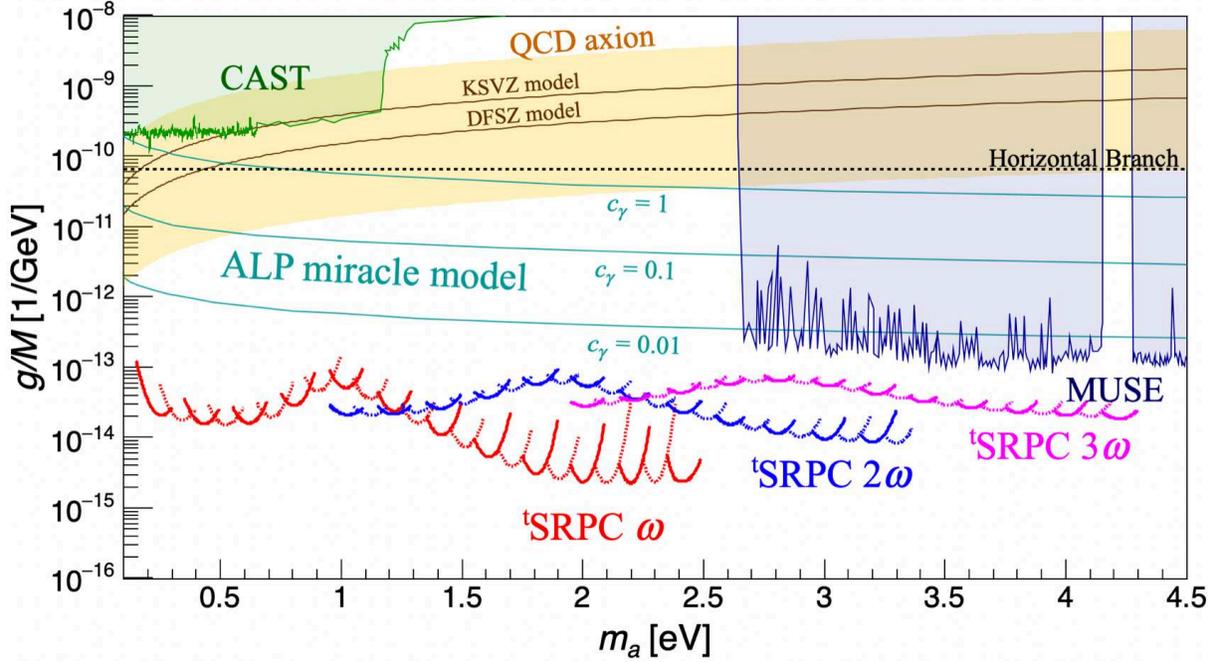}
\end{center}
\caption{
Expected sensitivities in the coupling-mass relation
for the pseudoscalar field exchange at a 95\% confidence level
by a three-beam stimulated resonant photon collider (${}^\mathrm{t}$SRPC) 
with focused short-pulsed lasers based on the beam parameters in Table \ref{Tab1}.
}
\label{Fig5}
\end{figure}

Based on parameters in Table \ref{Tab1},
Fig.\ref{Fig5} shows the reachable sensitivities in the coupling-mass relation
for the pseudoscalar field exchange at a 95\% confidence level by ${}^\mathrm{t}$SRPC.
The red, blue, and magenta solid/dashed curves show the expected upper limits
by ${}^\mathrm{t}$SRPC when we assume $\omega_c = 800$~nm (fundamental wavelength $\omega$), 
$400$~nm (second harmonic $2\omega$) and $267$~nm (third harmonic $3\omega$), respectively.
The ALP mass scanning is assumed to be performed with the step of $0.1$~eV.
Thanks to energy and momentum fluctuations at around the focal point, the same order sensitivities
are maintained within the assumed scanning step (the local minima of the parabolic behavior 
in the coupling correspond to different incident angle setups in Fig.\ref{Fig5}). 
For easy viewing, the solid and dashed curves are drawn alternatively.
These assumed photon sources are all available within the current technology~\cite{ELI}
in terms of the photon wavelength and energy per pulse.

These sensitivity curves are obtained based on the following condition.
In this virtual search, the null hypothesis is supposed to be fluctuations
on the number of photon-like signals following a Gaussian distribution
whose expectation value, $\mu$, is zero for the given total number of collision statistics.
The photon-like signals implies a situation where photons-like peaks are counted by a
peak finder based on digitized waveform data from a photodevice~\cite{SAPPHIRES00},
where electrical fluctuations around the baseline of a waveform cause 
both positive and negative numbers of photon-like signals. 
In order to exclude this null hypothesis a confidence level $1-\alpha$ is introduced as
\beq
1-\alpha = \frac{1}{\sqrt{2\pi}\sigma}\int^{\mu+\delta}_{\mu-\delta}
e^{-(x-\mu)^2/(2\sigma^2)} dx = \mbox{erf}\left(\frac{\delta}{\sqrt 2 \sigma}\right),
\eeq
where $\mu$ is the expected value of an estimator $x$ following the hypothesis, and
$\sigma$ is one standard deviation.
In this search, the estimator $x$ corresponds to the number of signal photons $N_S$ and
we assume the detector-acceptance-uncorrected uncertainty $\delta N_{S}$
as the one standard deviation $\sigma$ around the mean value $\mu=0$.
For setting a confidence level of 95\%, $2 \alpha = 0.05$ with $\delta = 2.24 \sigma$ is used, 
where a one-sided upper limit by excluding above $x+\delta$~\cite{PDGstatistics} is considered.
For a set of experimental parameters $P$ in Table~\ref{Tab1},
the upper limits on the coupling--mass relation, $m_a$ vs. $g/M$,
are then estimated by numerically solving the following equation
\beq
N_{obs} = 2.24 \delta N_S = {\cal Y}_{c+i}(m_a, g/M ; P) N_{shots} \epsilon.
\eeq

The horizontal dotted line shows the upper limit from the Horizontal Branch (HB) observation~\cite{HB}.
The purple area shows bounds by the optical MUSE-faint survey~\cite{MUSE}.
The green area is excluded by the helioscope experiment CAST~\cite{CAST}.
The yellow band shows the QCD axion benchmark models
with $0.07<|E/N-1.95|<7$ where KSVZ($E/N=0$)~\cite{KSVZ} and DFSZ($E/N=8/3$)~\cite{DFSZ}
are shown with the brawn lines.
The cyan lines show predictions from the ALP {\it miracle} model~\cite{MIRACLE}
with its intrinsic model parameters $c_{\gamma}=1.0, 0.1, 0.01$, respectively.

\section{Conclusion}
We have evaluated expected sensitivities to axion-like particles coupling to photons
based on the concept of a three-beam stimulated resonant photon collider with focused short-pulse lasers.
Within the current high-intensity laser technology reaching the pulse energy 1~J,
we found that the searching method can probe ALPs in the eV mass range 
down to $g/M = {\mathcal O}(10^{-14})$~GeV${}^{-1}$.
This sensitivity is sufficient to test the unexplored domain motivated by
the {\it miracle} model as well as the benchmark QCD axion models.

\section*{Acknowledgments}
K. Homma acknowledges the support of the Collaborative Research
Program of the Institute for Chemical Research at Kyoto University
(Grant Nos.\ 2018--83, 2019--72, 2020--85, 2021--88, and 2022--101)
and Grants-in-Aid for Scientific Research
Nos.\ 17H02897, 18H04354, 19K21880, and 21H04474 from the Ministry of Education,
Culture, Sports, Science and Technology (MEXT) of Japan.
Y. Kirita acknowledges support from the JST, the establishment of university fellowships for the creation of science technology innovation, Grant No. JPMJFS2129, and a Grant-in-Aid for JSPS fellows No. 22J13756 from the Ministry of Education, Culture, Sports, Science and Technology (MEXT) of Japan.


\begin{thebibliography}{99}
\bibitem{PQ}
R. D. Peccei and H. R. Quinn, Phys. Rev. Lett {\bf 38}, 1440 (1977)
\bibitem{AXION}
S. Weinberg, Phys. Rev. Lett {\bf 40}, 223 (1978);
F. Wilczek, Phys. Rev. Lett {\bf 40}, 271 (1978);
J. E. Kim, Phys. Rev. Lett. {\bf 43}, 103 (1979);
M. A. Shifman, A. I. Vainshtein and V. I. Zakharov, Nucl. Phys. B {\bf 166}, 493 (1980).
\bibitem{MIRACLE}
Ryuji Daido et al., "The ALP miracle revisited", Journal of High Energy Physics, 02 (2018) 104, 16th February 2018.
\bibitem{DEptp}
Y. Fujii and K. Homma, Prog. Theor. Phys {\bf 126}, 531 (2011); Prog. Theor. Exp. Phys. 089203 (2014) [erratum].
\bibitem{PTEP2014}
K. Homma, T. Hasebe, and K.Kume, Prog. Theor. Exp. Phys. 083C01 (2014).
\bibitem{PTEP2015}
T. Hasebe, K. Homma, Y. Nakamiya, K. Matsuura, K. Otani, M. Hashida, S. Inoue, S. Sakabe, Prog. Theo. Exp. Phys. 073C01 (2015).
\bibitem{PTEP2020}
A. Nobuhiro, Y. Hirahara, K. Homma, Y. Kirita, T. Ozaki, Y. Nakamiya, M. Hashida, S. Inoue, and S. Sakabe, Prog. Theo. Exp. Phys. 073C01 (2020).
\bibitem{SAPPHIRES00}
K. Homma, Y. Kirita, M. Hashida, Y. Hirahara, S. Inoue, F. Ishibashi, Y. Nakamiya, L. Neagu, A. Nobuhiro, T. Ozaki, M. Rosu, S. Sakabe and O. Tesileanu (SAPPHIRES), Journal of High Energy Physics, 12, (2021) 108.
\bibitem{SAPPHIRES01}
Y. Kirita, T. Hasada, M. Hashida, Y. Hirahara, K. Homma, S. Inoue, F. Ishibashi, Y. Nakamiya, L. Neagu, A. Nobuhiro, T. Ozaki, M. Rosu, S. Sakabe and O. Tesileanu (SAPPHIRES), Journal of High Energy Physics, 10, (2022) 176.
\bibitem{JHEP}
Kensuke Homma and Yuri Kirita, Journal of High Energy Physics, 09, (2020) 095.
\bibitem{UNIVERSE00}
Kensuke Homma, Yuri Kirita, Fumiya Ishibashi, Universe 7 (2021) 12, 479.
\bibitem{BJ}
J. D. Bjorken and S. D. Drell, {\it Relativistic Quantum Mechanicsh},
McGraw-Hill, Inc. (1964);
See also Eq.(3.80) in W. Greiner and J. Reinhardt, 
{\it Quantum Electrodynamics Second Edition}, Springer (1994).
\bibitem{Yariv}
Amnon Yariv, {\it Optical Electronics in Modern Communications} Oxford University Press (1997).
\bibitem{ELI}
https://eli-laser.eu
\bibitem{PDGstatistics}
See Eq.(36.56) in J. Beringer {\it et al.} (Particle Data Group), Phys. Rev. D {\bf 86}, 010001 (2012).
\bibitem{HB}
A. Ayalaet al., Phys. Rev. Lett. 113, 19, 191302 (2014).
\bibitem{MUSE}
Marco Regis, Marco Taoso, Daniel Vaz, Jarle Brinchmann, Sebastiaan L. Zoutendijk, Nicolas F. Bouch\'{e}, Matthias Steinmetz,
Physics Letters B, 814 (2021) 136075.
\bibitem{CAST} 
V. Anastassopoulos et al. (CAST), Nature Phys. {\bf13}, 584 (2017).
\bibitem{KSVZ}
J. E. Kim, Phys. Rev. Lett. 43, 103 (1979);
M. Shifman, A. Vainshtein, and V. Zakharov, Nucl. Phys. B166, 493 (1980).
\bibitem{DFSZ}
M. Dine, W. Fischler, and M. Srednicki, Phys. Lett. 104B, 199 (1981);
A. Zhitnitskii, Sov. J. Nucl. Phys. 31, 260 (1980).
\end{thebibliography}
\end{document}